\documentclass[10pt,twocolumn,letterpaper]{article}
\pdfoutput=1

\usepackage[pagenumbers]{cvpr} 

\usepackage{graphicx}
\usepackage{amsmath}
\usepackage{amssymb}
\usepackage{booktabs}

\usepackage{hyperref}
\hypersetup{pagebackref,breaklinks,colorlinks}

\usepackage[capitalize]{cleveref}
\crefname{section}{Sec.}{Secs.}
\Crefname{section}{Section}{Sections}
\Crefname{table}{Table}{Tables}
\crefname{table}{Tab.}{Tabs.}

\newcommand{\imaOne}{I} 
\newcommand{\imaTwo}{J} 
\newcommand{\FouOne}{\mathcal{F}} 
\newcommand{\FouTwo}{\mathcal{G}} 

\newcommand{\MSE}{\mathrm{MSE}} 
\newcommand{\FRC}{\mathrm{FRC}} 
\newcommand{\FRCloss}{\mathcal{L}_\mathrm{FRC}} 
\newcommand{\Lone}{\mathrm{L_1}} 
\newcommand{\Ltwo}{\mathrm{L_2}} 
\newcommand{\Loneloss}{\mathcal{L}_1} 
\newcommand{\Ltwoloss}{\mathcal{L}_2} 

\title{\LARGE \bf
Image quality measurements and denoising using \\ Fourier Ring Correlations
}

\author{J. Kaczmar-Michalska$^{\ast \dagger\ddagger}$, N. R. Hajizadeh$^{\ast}$, A.J. Rzepiela$^{\ast  \diamond}$ and S.F. N\o{}rrelykke$^{\ast  \diamond}$\vspace{0.3cm}\\ 
$^{\diamond}$Correspondence to andrzejr@ethz.ch and norsimon@ethz.ch\\
$^{\ast}$Scientific Center for Optical and Electron Microscopy (ScopeM),\\ 
Eidgen\"ossische Technische Hohschule Z\"urich, (ETH Zurich), 8093 Z\"urich, Switzerland\\
$^{\dagger}$TOOPLOOX, 53-601 Wroc\l{}aw, Poland\\
$^{\ddagger}$Wrocław University of Science and Technology, 50-370 Wroc\l{}aw, Poland
}

\begin{document}

\maketitle

\begin{abstract}
Image quality is a nebulous concept with different meanings to different people. To quantify image quality a relative difference is typically calculated between a corrupted image and a ground truth image. But what metric should we use for measuring this difference? Ideally, the metric should perform well for both natural and scientific images. The structural similarity index (SSIM) is a good measure for how humans perceive image similarities, but is not sensitive to differences that are scientifically meaningful in microscopy. In electron and super-resolution microscopy, the Fourier Ring Correlation (FRC) is often used, but is little known outside of these fields. Here we show that the FRC can equally well be applied to natural images, e.g.\ the Google Open Images dataset. We then define a loss function based on the FRC, show that it is analytically differentiable, and use it to train a U-net for denoising of images. This FRC-based loss function allows the network to train faster and achieve similar or better results than when using $\Lone$- or $\Ltwo$-based losses. 
We also investigate the properties and limitations of neural network denoising with the FRC analysis.
\end{abstract}

\section{\textbf{INTRODUCTION}}

\subsection{\textbf{Neural networks for image denoising}}

One of the most fundamental problems in image processing is that of image denoising, formulated as recovering a high quality image from its noisy (degraded) realisation.
Denoising methods can be grouped into traditional methods---including spatial and frequency based filtering as well as wavelet transform-based methods \cite{BM3DNetAC}, BM3D being a well known example \cite{dabov2009bm3d}---and the more recent machine learning based approaches, mainly the broad family of convolutional neural network (CNN) models.

Traditional methods often rely on prior knowledge and explicit assumptions about the character of the corruption, whereas learning-based approaches are more data-driven and learn to differentiate the signal from the noise by seeing a large number of examples. Following the rapid development of deep learning over the past decade, there have been a multitude of deep learning denoising models proposed, for instance DnCNN \cite{zhang2017beyond}, FFDNet for blind denoising (when the noise level is unknown) \cite{zhang2018ffdnet}, BRDNET \cite{tian2020image}, and U-NET based models \cite{weigert2018content} to name but a few. Over the last few years, several deep learning approaches were introduced that eliminated the need for ground truth data (GT), with the most prominent being Noise2Noise (N2N) \cite{lehtinen2018noise2noise} which recovers signals under different types of corruptions, from pairs of corrupted images, if the expectation of the corrupted data is the same as the clean target. N2N gives state of the art performance \cite{batson2019noise2self}, but requires a setup where a target is measured a several times (independent noise realizations).

\begin{figure}[tbp]
\centering
\includegraphics[width=1.0\columnwidth]{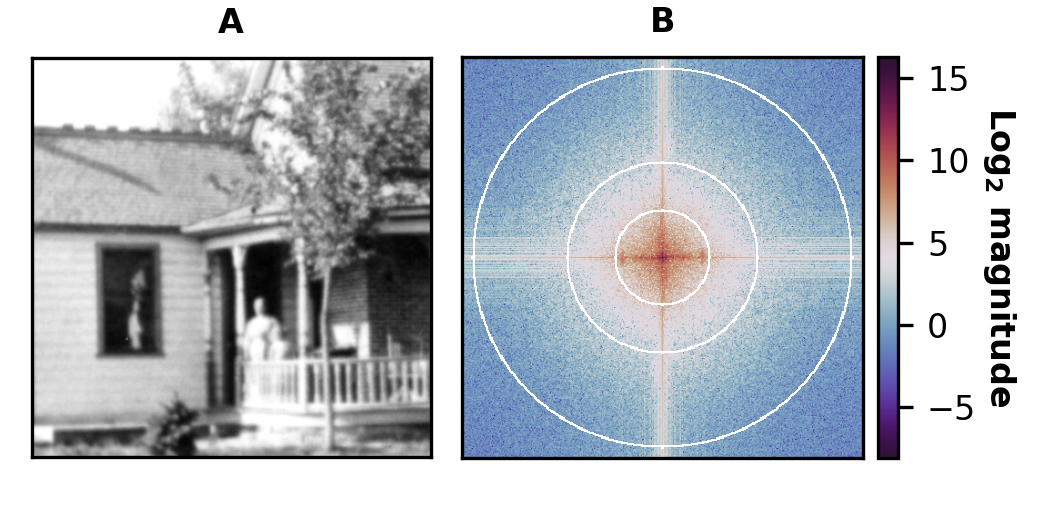}
\caption{$\mathbf{A}$, an image and $\mathbf{B}$, $\log_2$ magnitudes of its power spectral density in 2D Fourier space with three examples of concentric rings (white), as used to calculate the Fourier ring correlation (FRC) values. 
The low frequency information with high magnitudes is in the center of the transformed image.}
\label{frc_explanation}
\end{figure}

More recently, the focus has shifted towards improving denoising with only a single corrupted image as input, and several networks such as Noise2Void \cite{krull2019noise2void} or Noise2Self (N2S) \cite{batson2019noise2self} have been proposed. 
During training, these methods systematically remove/mask parts of the input, e.g.\ the center pixel,  and infer the pixel value from just its neighbourhood. In probabilistic unsupervised denoising, a noise model is incorporated in the process, in order to refine the posterior distribution of the signal. The noise model can be created either non-parametrically through sampling examples \cite{krull2019probabilistic,prakash2020fully}, or by bootstrapping existing data and formulating the noise model, for example as a Gaussian mixture model (GMM) or Gaussian/Poisson noise model \cite{laine2019high}. These denoisers outperform their simpler counterparts (N2S) and achieve performance comparable to training schemes with the ground truth available. Last, but not least, the DivNoising network explicitly samples from an approximate posterior of possible true signals obtained by using a variational autoencoder and generates an ensemble of possible predictions for the denoised image  \cite{prakash2021fully}. This approach both highlights and explicitly addresses the inherent ambiguity in the denoising problem.

\subsection{\textbf{Denoising metrics}}

A concept directly relevant to image denoising, and in particular learning based approaches, is the score or metric. Here we use "metric" not in the mathematical sense of a distance function, but simply to refer to the function used to calculate the difference between two images or the quality or resolution of a single image;  we use "loss" to refer to the function used to train a network. The metric and the loss can be the same function.

The most widely used metric and loss for image denoising with neural networks is the mean squared error (MSE) \cite{zhao2015loss}, also called the $\Ltwoloss$ loss, which is based on the pixel-wise difference between two images. If $\imaOne$ and $\imaTwo$ are monochrome images of size $M\times N$, then the MSE is defined as follows 

\begin{equation}
\Ltwoloss(\imaOne,\imaTwo) \equiv \MSE 
= \frac{1}{MN}\sum_{m=0}^{M-1}\sum_{n=0}^{N-1}\left [ \imaOne_{m,n} - \imaTwo_{m,n} \right ]^{2} \enspace ,
\end{equation}
where $m,n$ are discrete pixel coordinates.
It is therefore a representation of absolute error, where values closer to zero indicate a higher quality. The use of MSE assumes that the impact of noise is independent of the local characteristics of the image, and implicitly assumes a Gaussian noise model. 
A related metric for image quality assessment is the peak-signal to noise ratio (PSNR), which is defined in terms of the MSE as

\begin{equation}
\mathrm{PSNR(\imaOne,\imaTwo)} \equiv 10 \log_{10}\left ( \frac{\max^2(\imaOne)}{\MSE(\imaOne,\imaTwo)} \right ) \enspace,
\end{equation}
where $\max(\imaOne)$ is the maximum (the peak) \textit{possible} value of the clean (signal) image, e.g.\ 255 for an 8-bit image, the MSE is a measure of the noise that separates $\imaOne$ and $\imaTwo$,  and the ratio is measured in decibel. Both the MSE and PSNR are straightforward to determine and are physically intuitive as metrics. However, they often do not correlate well with perceived image quality \cite{wang2003multiscale}. Complementary approaches for assessing image quality exist, such as the structural similarity index (SSIM) \cite{wang2004image}, and more recently its multi-scale variant (MS-SSIM) \cite{wang2003multiscale}, with the latter being the most successful image metric in terms of mimicking human judgement. However, in the context of quantifying signal from images, e.g.\ scientific images, mimicking human judgement is not the absolute priority.

In the field of electron microscopy, where high voltage electrons are used to image materials, a metric based on the correlation of images in Fourier space, the Fourier Ring Correlation (FRC) \cite{vanHeel1982,SAXTON:1982jx} and its 3D variant Fourier Shell Correlation (FSC) \cite{harauz1986exact}, has been in use since their introduction in the 80ies. The FRC's original intended use was to estimate resolution from images, and their 3D models, of proteins acquired by single particle cryo-electron microscopy \cite{van2005fourier}. The FRC and some closely related measures have, however,  also found recent application in other fields, for example in the fluorescence  and super-resolution microscopy communities, where they are used for image restoration \cite{koho2019fourier} and reconstruction \cite{Berberich2021}, measuring resolution \cite{BANTERLE2013363,Descloux:2019kh}, and detecting optical artifacts \cite{Culley:2018hx}.

The FRC/FSC is based on a normalized cross-correlation  between two images, calculated in the frequency domain. The calculation starts with dividing the spatial frequency spectra of the two images into a series of concentric rings, see Figure \ref{frc_explanation}.  The FRC value, a scalar, of each ring is calculated like this

\begin{equation}
  \label{eq:FRC}
  \FRC(\imaOne,\imaTwo;r) \equiv \frac{ \sum_S \FouOne_{p,q} \, \FouTwo^*_{p,q} }{ \sqrt{  \sum_S |\FouOne_{p,q}|^2  \;  \sum_S |\FouTwo_{p,q}|^2} } 
\end{equation}
where $S$ is the set of coordinates where $p^2+q^2=r^2$, i.e., a ring of radius $r$ in Fourier space, $^*$ denotes complex conjugation, $\FouOne$ is the discrete 2D Fourier transforms of image $\imaOne$ 
\begin{equation} \label{eq:Fourier}
    \FouOne_{p,q} \equiv \tilde{\imaOne}_{p,q} = 
    \sum_{m=0}^{M-1} \sum_{n=0}^{N-1} \imaOne_{m,n} \, e^{-i2\pi (pm/M + qn/N)} \enspace , 
\end{equation}
and an analogous expression holds for $\FouTwo$, the transform of $\imaTwo.$
The sum in the numerator of the FRC is real, as explained in \cite{van1987similarity}, since Fourier transforms of images exhibit Friedel symmetry and pixels placed on opposite sides of the frequency ring form conjugate pairs. 
The maximal number of complete rings for a $N\times N$ digital image is $\frac{N}{2}$, following the Nyquist frequency sampling criterion. 
Higher frequency rings, up to $\frac{N}{\sqrt{2}}$, are also partially present. 
The FRC takes on values between $+1$, the value found when comparing two identical images (perfect correlation), and $-1$ (anti-correlation) for comparing an image and its inverted copy. 
For noisy images the lower limit is $0$ (uncorrelated), when e.g.\ noise is compared to a ground truth image. 
In practice, the FRC is often calculated in non-overlapping annuli of thickness larger than one, i.e.\ a larger band of frequencies are included for each FRC value, mostly to suppress fluctuations.
Some useful mathematical properties and limit behaviors of the FRC are derived in section \ref{theory}. 

\subsection{\textbf{Aim and approach}}
Here we build on the results of \cite{koho2019fourier} in terms of translating the use of FRC as a resolution assessment to also be used in the wider context of image denoising and image quality assessment. Our contributions are three-fold. 

First, we derive some mathematical facts of the FRC: Its analytic differentiability, limit properties, and invariant behaviour in a number of common scenarios, such as image-scaling, additive noise, and spatial filtering.

Next, we show that the FRC can be used for assessing image quality, more specifically we show that it constitutes a highly informative quality metric for natural images with performance on par with SSIM, but with the additional advantage of providing the quality score as a  function of signal frequency. Previous works have highlighted the importance of combining numerical and graphical measures to judge image quality, and that a single scalar value is not sufficient to comprehensively assess the range of degradations \cite{eskicioglu1995image}. 

Then, we showcase how the FRC can be used to define a loss function for neural network training, and provide its application for denoising tasks on different noise types. Networks trained with the FRC-based loss, train faster due to its gradient shape, and can denoise several types of degradations.  

Finally, with the help of the FRC metric, we make observations on the characteristics of image denoising. In section \ref{spectrum} we show how denoising performance correlates with the power spectrum of the images, i.e.\ exhibiting greater performance on frequencies with higher signal. 

\section{\textbf{Theory}}\label{theory}

\subsection{\bf{Differentiability of the FRC-loss}} \label{FRC_diff}

Having defined the FRC in Eq.~(\ref{eq:FRC}) we introduce the FRC-loss as the sum over all annuli:
\begin{equation}
  \label{eq:FRC-loss}
  \FRCloss(\imaOne,\imaTwo) \equiv 1- \sum_r \FRC(\imaOne,\imaTwo;r)
\end{equation}

Consider the FRC-loss $\FRCloss$ as defined in Eq.~(\ref{eq:FRC-loss}) but rewritten, simplifying notation by suppressing dependence on $\imaOne$ and $\imaTwo$, as
\begin{equation}
  \FRCloss = 1- \sum_r \frac{ A(r) }{ B(r) }  
\end{equation}
with
\begin{align}
  A(r)  &= \sum_S \FouOne_{p,q} \, \FouTwo^*_{p,q} \\
  B(r)  &=  \sqrt{  \sum_S |\FouOne_{p,q}|^2  \;  \sum_S |\FouTwo_{p,q}|^2} \enspace .
\end{align}

Now, $\FRCloss$ is a sum of terms (plus a constant), so if each term is differentiable, then so is the sum, i.e., we can focus on the differentiability of the FRC itself.
This again reduces to three steps, namely to showing that $A$, $B$, and $A/B$ are differentiable.
Writing $X'$ for the pixel-wise differentiation $\partial X / \partial \imaOne_{k,l}$, we have:
\begin{equation}
  \FRC'(r) = \frac{A'B-AB'}{B^2}  
\end{equation}
and
\begin{equation}
  B'(r) = \sqrt{ \frac{ \sum_S |\FouTwo_{p,q}|^2 }{ \sum_S|\FouOne_{p,q}|^2 } } \,
  \sum_S \FouOne_{p,q} \FouOne'_{p,q}
\end{equation}
and 
\begin{equation}
  A'(r) =   \sum_S  \FouTwo^*_{p,q} \, \FouOne'_{p,q}
\end{equation}
with
\begin{equation}
  \FouOne'_{p,q}=  e^{-i2\pi (pk/M + ql/N)} \enspace .
\end{equation}
Here, we used that $\FouTwo$ is independent of $\imaOne$, that $\partial \imaOne_{m,n} / \partial \imaOne_{k,l} = \delta_{m,k} \, \delta_{n,l}$, and the definition of the Fourier transform Eq.~(\ref{eq:Fourier}).
All of the these terms are well defined as long as none of the images are empty, meaning that $\FRCloss$ is differentiable and hence mathematically valid as a loss function for training-schemes that update weights and biases in neural networks via back-propagation.
\subsection{\bf{FRC limits}} \label{FRC_limits}
\subsubsection{Large-frequency FRC approximation}
\label{sec:large-freq-app}
Let us explore how the FRC behaves when one image $\imaOne$ is taken as ground truth and the other is a copy with noise added $\imaTwo = \imaOne + \eta$, with the expectation values $\left< \eta \right> = 0$ and $\left< \eta_{m,n} \, \eta_{m',n'} \right> = \sigma_\eta^2 \, \delta_{m,m'} \, \delta_{n,n'}$. 
The linearity of the Fourier transform and the characteristics of the noise directly gives
\begin{equation}
    \FouTwo_{p,q} = \FouOne_{p,q} + \tilde{\eta}_{p,q}
\end{equation}
and the numerator of the FRC becomes
\begin{equation}
  \label{eq:FRC_num}
  \sum_S  \FouOne \, \FouTwo^* = 
  \sum_S \FouOne \, [\FouOne + \tilde{\eta}]^* 
  \simeq n_r \left< |\FouOne|^2 \right>
  \simeq \sum_S |\FouOne|^2
\end{equation}
where we approximated the sum over the annulus in frequency space with the expectation value, which is zero for the Fourier transformed noise $\left<\tilde{\eta}_{p,q}\right> = 0$, and where $n_r$ is the number of elements in S, i.e.\ number of pixels in frequency space satisfying $p^2 + q^2 = r^2$. 
This result depends on $r$, but not on the strength of the noise, and the approximation should be better the larger the area of the annulus, i.e.\ for higher frequencies.

For the, squared, denominator we similarly find
\begin{equation}
  \label{eq:FRC_den}
  \sum_S  | \FouOne |^2 \,  \sum_S | \FouTwo |^2 \simeq
  n_r^2 \left< |\FouOne|^2 \right> \left[ \left< |\FouOne|^2 \right> + N M\,\sigma^2_\eta \right] \enspace .
\end{equation}

Forming the ratio of these two approximations and expanding to first order in the noise $\sigma^2$, we get
\begin{align}
  \FRC(\imaOne, \imaOne + \eta;r) &\simeq  \frac{1}{\sqrt{1 + NM \, \sigma_\eta^2 / \left<|\FouOne|^2\right>} } \label{eq:FRC_highFreq} \\ 
  & \simeq 1 - \frac{1}{2} NM \, \sigma_\eta^2 / \left< |\FouOne|^2\right>  \label{eq:FRC_Taylor} 
\end{align}
which, as the noise goes to zero, approaches unity.

If, additionally, we assume that the power spectral density of the image scales as one over frequency $\left<|\FouOne|^2\right> \propto 1/r$, then we see from Eq.~(\ref{eq:FRC_highFreq}) that the FRC will take on the characteristic form
\begin{equation}
    \label{eq:16}
    \FRC(\imaOne, \imaOne + \eta;r) \simeq \frac{1}{\sqrt{1+ar^2}}
\end{equation}
where $a$ is a factor that contains the frequency independent amplitudes of the noise and the image.
That is, the FRC decays smoothly from one to zero as a function of frequency $r$, with an inflection point around $1/\sqrt{2a}$.

\subsubsection{High noise, large frequency FRC approximation}
\label{sec:high-noise-large}
For high noise, but still in the approximation of large frequencies, the FRC in Eq.~(\ref{eq:FRC_highFreq}) trivially becomes 
\begin{equation}
  \label{eq:9}
  \FRC(\imaOne, \imaOne + \eta;r) \simeq \frac{ \left<|\FouOne|^2\right>  }{ NM \,\sigma_\eta^2 } 
  \simeq \frac{ \sum_S |\FouOne|^2  }{ n_r N M \,\sigma_\eta^2 } \enspace .
\end{equation}
That is, the FRC is inversely proportional to the strength of the noise in this limit and approaches zero as the noise grows.
Here, by large noise, we mean $NM\sigma^2_\eta \gg \left<|\FouOne|^2\right>$.

If, like above, we assume that the power of the signal scales as one over frequency, we see that the FRC will simply exhibit that same behavior, as can also be gleaned from Eq.~(\ref{eq:FRC_highFreq}) in the limit of large noise:
\begin{equation}
    \label{eq:18}
    \FRC(\imaOne, \imaOne + \eta;r) \simeq \frac{1}{r\sqrt{a}}
\end{equation}

\subsection{\bf{Mathematical invariances of the FRC}} \label{math_prop}
\subsubsection{Invariance to image scaling}
Scaling of the images by multiplicative factors $a_1>0$ and $a_2>0$ has no effect on the FRC:
\begin{equation}
    \FRC(a_1 \imaOne, a_2 \imaTwo; r) =     \FRC(\imaOne, \imaTwo; r) \enspace .
\end{equation}
This is seen by inspection of the definition of the FRC and by remembering the linearity of the Fourier transform: The Fourier transformed images appears with the same power in the numerator and denominator, so the factors $a_1$ and $a_2$ cancel out. If we choose a negative scaling factor, the sign of the FRC flips. 

\subsubsection{Pseudo-invariance to image offsets}
Adding constants $b_1$  and $b_2$ to an image only affects the value of the FRC at zero frequency 
\begin{align}
    \FRC(b_1 + \imaOne, b_2 + \imaTwo; r>0) & =     \FRC(\imaOne, \imaTwo; r>0) \\
    \FRC(b_1 + \imaOne, b_2 + \imaTwo; r=0) & \neq     \FRC(\imaOne, \imaTwo; r=0)
\end{align}
This, again follows directly from the properties of the Fourier transform and the definition of the FRC.

\subsubsection{Invariance to image filtering}
If an image is filtered in the Fourier domain by an $r$-dependent function that is radially symmetric, real, and positive, the FRC doesn't change.
This is easily seen, as such a filter would simply show up as a multiplicative, $r$-dependent factor (a real and positive scalar), in both the numerator and the denominator of the FRC, and thus cancel out. This instantly tells us, via application of the convolution theorem, that a certain class of \textit{spatial} filters also will not change the FRC. Any spatial filter whose Fourier transform is real, positive, and only depend on $r$ will leave the FRC invariant. A classic example would be smoothing in the image-domain by convolution with an isotropic 2D Gaussian function $g$ (the Fourier transform of a Gaussian is a Gaussian), which will have no effect on the FRC
\begin{equation} \label{eq:FRC_gauss}
    \FRC(\imaOne, g\star\imaTwo; r)  = \FRC(\imaOne, \imaTwo; r) \enspace,
\end{equation}
where $\star$ denotes convolution.

\section{\textbf{Results}}
\subsection{\textbf{FRC as an Image Quality Metric}}

\begin{figure}[tbp]
\centering
\includegraphics[width=1.0\columnwidth]{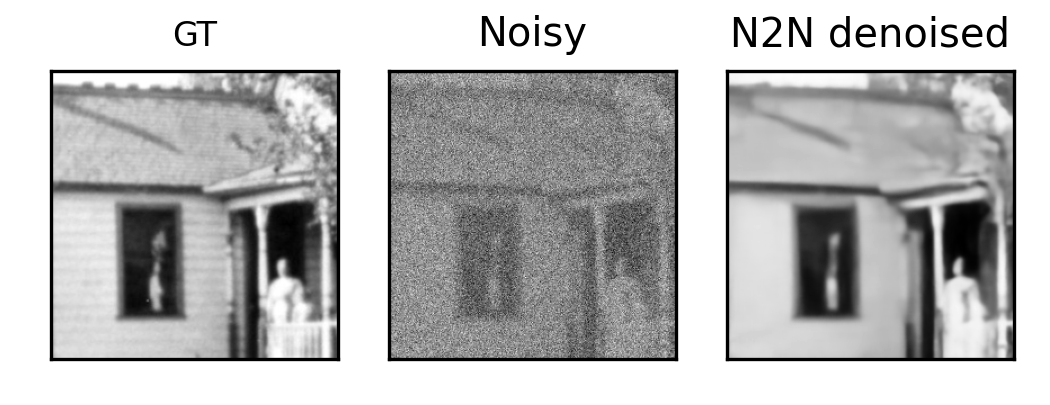}
\centering
\includegraphics[width=1.0\columnwidth]{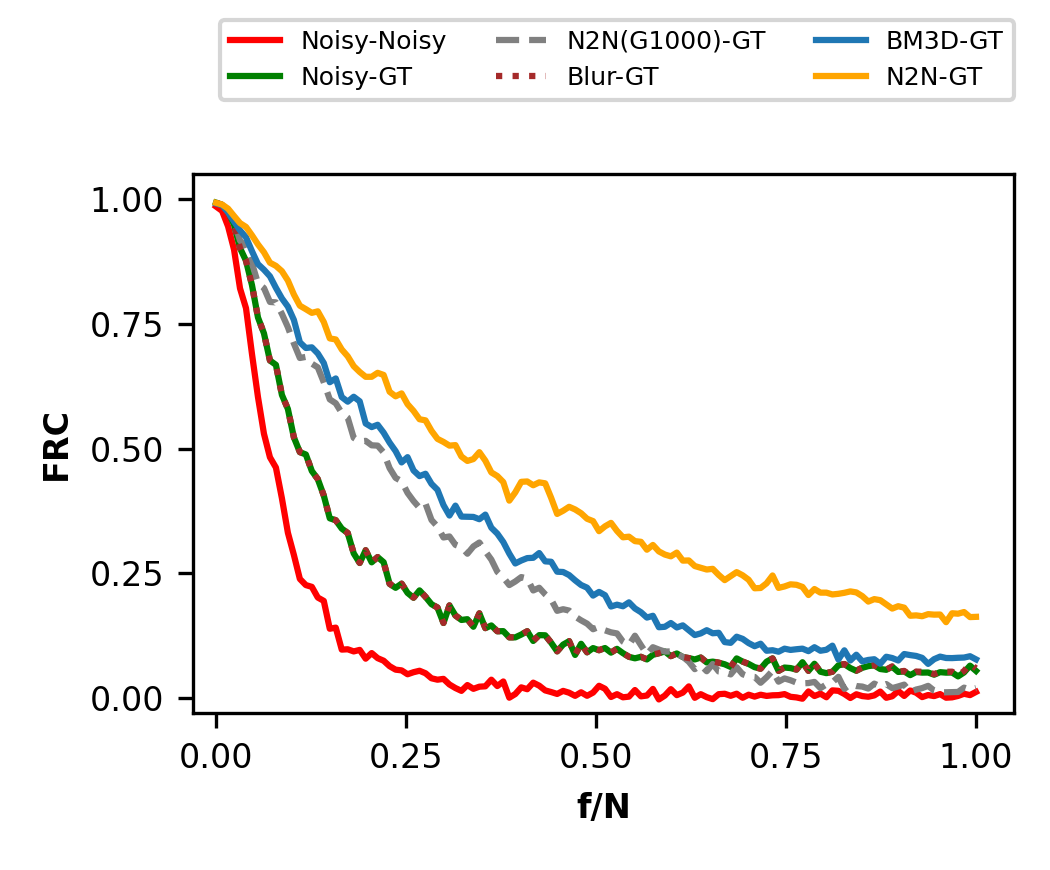}
\caption{N2N model performs best, Gaussian filter has no effect. To assess denoising performance, the FRC can be calculated between the denoised image and the ground truth or the noisy image. Higher FRC values indicates better restoration. Several denoising approaches are shown (N2N-GOI network vs.\ GT, orange line; BM3D method vs.\ GT, blue line; Gaussian filter vs.\ GT, brown dotted line; N2N-GOI1000 network vs.\ GT, grey dashed line) along with curves showing FRC for noisy data (noisy vs.\ noisy image, red line; noisy vs.\ GT image, green line). Each curve shows an average FRC for 50 natural test images from the GOI set, pre-processed with a Hann window to suppress edge artifacts in the Fourier signal. $f/N$ denotes frequency normalised by the Nyquist frequency for the images.}
\label{frc_example}
\end{figure}

When FRC is determined between a noisy/restored image and the ground truth, the resulting curve provides information on how much the two images correlate over the frequency spectrum, as depicted in Figure \ref{frc_example}. 
The one-dimensional FRC view is advantageous as one can assess the  quality of the signal as a function of frequency, a distinct advantage when comparing to scalar metrics. 
However, the area under the curve  can also be calculated for each FRC curve, producing a scalar, $1- \FRCloss$, that can be compared to the more traditional scalar loss-measures ($\Loneloss$, $\Ltwoloss$, PSNR, SSIM). 

To provide an example of the metric in action, we compare several FRC curves for noisy and denoised images in Figure \ref{frc_example}. The FRC is calculated between each noisy/denoised image and its corresponding pair image (ground truth, noisy) for 50 test images in the Google Open Images (GOI) dataset v5 \cite{OpenImages} and averaged (see Supplementary Materials). 

The red line in Figure \ref{frc_example} shows that the FRC between two noisy realizations of the same signal has an overall low value, especially for higher frequencies, following the decay described in Eq.~(\ref{eq:18}). When comparing a ground truth (GT) image and one of the noisy realisations we observe a higher FRC over all frequencies. This is expected, since lower noise levels, or denoising, of one of the images should increase the FRC (see section \ref{FRC_limits}). The standard BM3D algorithm \cite{dabov2007image} restores some of the signal from the noisy image and improves the FRC curve. However, a simple Gaussian filter (with sigma 1)  only scales the mean power spectra, effectively dumping higher frequencies, but does not affect the FRC curve,  (brown dotted line, see also Eq.~(\ref{eq:FRC_gauss}) ). On the other hand, a U-net based N2N network, trained on 50,000 images from the GOI dataset (see Supplementary Materials), is a better denoiser than BM3D, with particularly good performance in the medium and high frequency range (orange line). When the denoising model is trained on a smaller training dataset (1000 images from the GOI set) a decrease in the performance of the denoiser is observed, as expected.

\subsection{\textbf{Comparison of FRC with standard metrics across corruptions}}

To investigate how well $\FRCloss$ captures distortions, in comparison to more established metrics such as MSE and SSIM, we compare the performance of these metrics using the KADID-10 dataset containing artificially distorted images that have corresponding differential mean opinion scores (DMOS). DMOS is a human perception image quality score (between 1 and 5), derived through the use of internet crowdsourcing, see \cite{kadid10k}). KADID-10 contains 5 distortion levels for each type of distortion. Here, we considered a set of 4 distortion types on the full set of 81 images from the KADID-10 dataset. For each image and distortion type, the three metrics ($1- \FRCloss$, MSE and SSIM) were applied to obtain an image quality estimation scalar, resulting in 5x3 estimations (five levels of distortion, three methods to evaluate). For each method, the five estimations were correlated with the DMOS by using Pearson's correlation coefficient (using all images in the dataset). Our results are summarized in Figure S1 and Table~\ref{table_corruption}. 

The performance of $1- \FRCloss$ is competitive (Gaussian and impulse noise) or better (jitter and motion blur) than SSIM and overall better than the MSE metric.  Both the SSIM and $1- \FRCloss$ metrics are sensitive to the quality of high frequency signals in images, which is also expressed in the DMOS, as discussed in \cite{wang2003multiscale}. In contrast, the MSE metric, as shown in Figure \ref{lossfnt_shape}, depends mostly on the low frequency characteristic of an image.

\begin{table}[tbh!]
\centering
\begin{tabular}{c| p{1.65cm} p{1.65cm} p{1.65cm}} 
 \hline
\hline
Distortion & $1-\mathbf{\FRCloss}$ & MSE & SSIM \\ [0.5ex] 

Metric &  &  &  \\ [0.5ex] 
 \hline\hline
 Gaussian & 0.94  (0.01) & 0.93  (0.01) & \bf{0.95  (0.01)}  \\
Impulse & \textbf{0.94  (0.01)} & 0.93  (0.01) & \bf{0.94  (0.01)} \\
Jitter & \bf{0.97  (0.01)} & 0.89  (0.02) & 0.95  (0.01)  \\
Motion blur &  \bf{0.96  (0.01)} & 0.95  (0.01) & 0.94  (0.01)  \\ [1ex] 
 \hline
\end{tabular}
\captionof{table}{Performance comparison of image quality assessment models on 81 images from the KADID-10 dataset. The performance is measured by calculating each metrics and correlating it to the differential mean opinion score, DMOS, obtained from the dataset. Average Pearson's  correlation coefficients for $\FRCloss$, MSE and SSIM vs.\ DMOS (over all 81 images in the data set) with the standard error are shown. The best performance, for each type of distortion, is shown in bold.}
\label{table_corruption}
\end{table}

\subsection{\textbf{FRC as a loss function in neural networks}}

We have now shown that the FRC can be used as a metric to assess frequency-dependent signal quality in natural images, with a sensitivity comparable to or super-seeding the more well-known SSIM metric. 
Next, we explore the characteristics of $\FRCloss$ as a loss function for training denoising neural networks. We begin by analysing the FRC loss function dependency on spatial signal frequency, followed by experiments measuring denoising performance across different noise corruptions. We compare the results to the standard $\Ltwoloss$ (MSE) and, where relevant, the $\Lone$ loss, $\Loneloss$. 

\subsubsection{\textbf{FRC loss dependency on signal spatial frequency}}
\begin{figure}[tbh!]
\centering
\includegraphics[width=1.0\columnwidth]{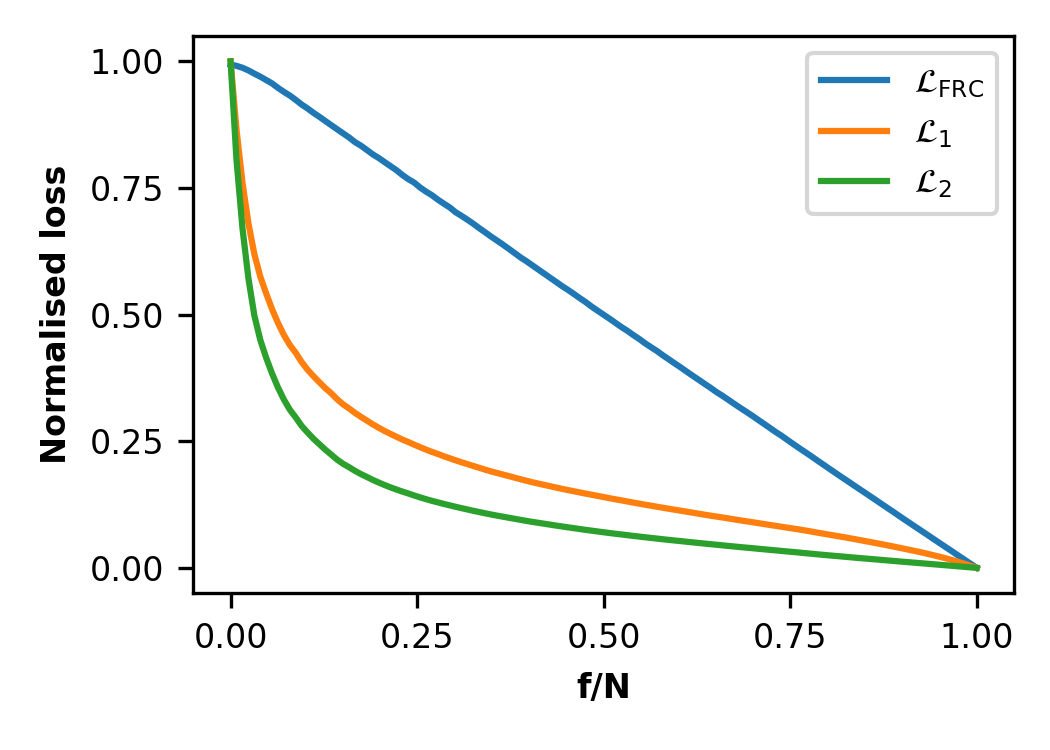}
\caption{$\Ltwoloss$ has weak sensitivity to image details. The plot shows dependency of $\Loneloss$, $\Ltwoloss$, and $\FRCloss$ to the signal from natural images as function of low-pass frequency. Iteratively, over the frequency from 1 up to Nyquist,  a cutoff  in Fourier space was applied to remove signal above a given frequency threshold. For each resulting low-pass filtered image $\Loneloss$, $\Ltwoloss$, and $\FRCloss$ were calculated relative to the ground truth image. Averages over 50 test GOI images are plotted. Before averaging, each image-loss curve was normalized. $f/N$ denotes frequency normalised by the Nyquist frequency for the images.}
\label{lossfnt_shape}
\end{figure}

To understand how the different losses restore information as a function of frequency we used  low-pass filtering for each of the 50 images in the test dataset. 
In Fourier space, a range of cutoffs between zero and the Nyquist frequency were applied to remove signal above a given frequency threshold. 
Afterwards, images were inverse Fourier transformed, and the three loss functions $\Loneloss$, $\Ltwoloss$, and $\FRCloss$ were calculated relative to the ground truth images. 
After processing all available frequencies, the loss curves were normalised, averaged for the full data-set, and  shown in Figure \ref{lossfnt_shape}. 
We observe that almost all of the change in $\Ltwo$ happens in a narrow frequency range close to zero, indicating a strong sensitivity to low spatial frequencies or, reversely, a low sensitivity to high-frequency signals. 
This may not be an issue for some applications, since most of the signal strength is contained in the low frequencies in natural images; for scientific images this is not universally the case though.. 
In contrast to this behavior $\FRCloss$ depends equally strongly on all frequencies as shown by its linear response to cut-off frequency. 
The $\Lone$-loss behaves similarly to $\Ltwoloss$, with slightly higher sensitivity to the high frequency features. 
These result suggests that a network with the $\Ltwo$-loss might be more difficult to train on high frequency  signal, when compared to models using $\Lone$ or $\FRCloss$.

\subsubsection{\textbf{FRC loss and noise distributions}}
\label{loss}

In the theory section \ref{FRC_diff} we showed  that $\FRCloss$ is differentiable and can be used as a loss-function in neural network model training. Here,  we test its performance in image denoising. We compare denoising with $\FRCloss$ against $\Loneloss$ and $\Ltwoloss$ based denoising for two artificial corruptions, Gaussian and log-normal noise, as well as experimental noise in single particle cryo-electron microscopy images. 

Since minimising $\Ltwoloss$ corresponds to maximising a model log-likelihood when denoising data with a Gaussian noise, $\Ltwoloss$ should give optimal results for such dataset. 
Similarly, $\Loneloss$ should be optimal for denoising images with log-normal noise. 
The distribution of noise in electron microscopy images can be approximated, but in principle it is not known, making a choice of a loss function more arbitrary. The properties of the applied noise distributions are given in the Supplementary Materials.

\begin{figure*}[h]
\vspace{1.5cm}
\centering
\includegraphics[width=1.0\textwidth]{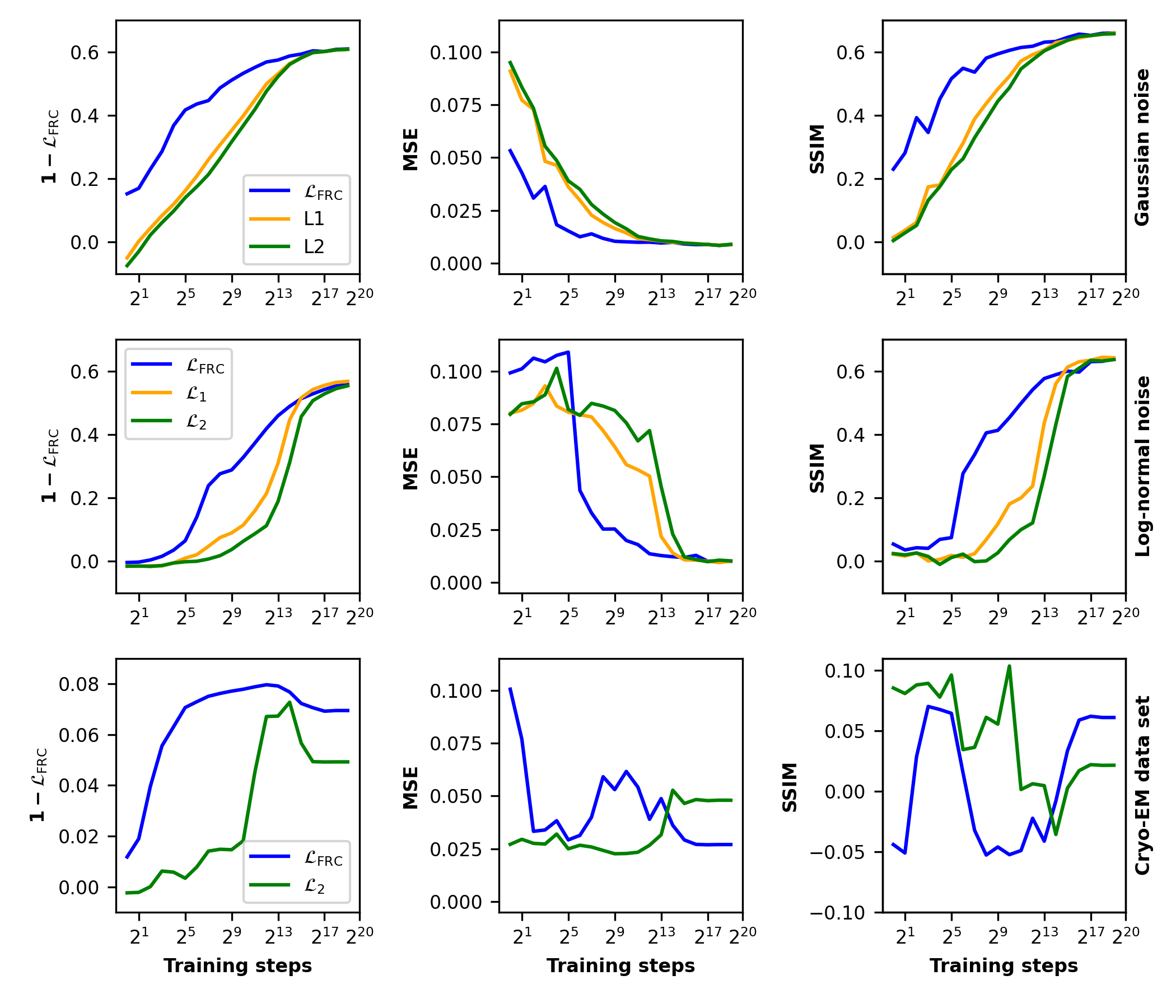}
\caption{Comparison of loss function performance in N2N denoising network trained for two artificial corruptions and a real-world noise. Top row: Gaussian noise (GOI data set), middle row: log-normal noise (GOI data set), bottom row: electron microscopy single particle data. 
$\FRCloss$ (blue line) was compared to the statistically best loss function for each corruption ($\Ltwoloss$, green line, for Gaussian noise, and $\Loneloss$, orange line, for log-normal noise). 
For electron microscopy images, results for $\FRCloss$ and $\Ltwoloss$ are plotted. 
$\FRCloss$, MSE and SSIM metrics averages for 50 (GOI) or 200 (cryoEM) test images, for the corresponding data sets, are shown.}
\label{loss_training}
\vspace{0.5cm}
\end{figure*}

Here, we test the hypothesis that $\FRCloss$ will provide comparable results to $\Ltwoloss$ for Gaussian noise, to $\Loneloss$ for log-normal noise, and possibly high quality results for the cryo-EM data images. In Figure \ref{loss_training} we see that the FRC-loss allows significantly faster convergence of network parameters, for all three data sets. 
Notably, SSIM values for Gaussian noise after $2^5$ training steps with $\FRCloss$ are comparable to SSIM values for $\Ltwoloss$ after $2^{10}$ training steps (top row, right column in Figure \ref{loss_training}). This is likely due to stronger gradients for high frequency features for this loss. 
We also observe that $\Lone$-loss for the log-normal noise data set converges faster than $\Ltwo$-loss, but slower than $\FRCloss$, which is consistent with the results shown in Figure~\ref{lossfnt_shape}. 
     
Eventually, all three loss functions converge to the same level: Final results from long training ($2^{19}$ steps $\simeq 200$ epochs) are similar for all three loss functions and the three metrics we used ($\FRCloss$, MSE, SSIM). In \cite{lehtinen2018noise2noise} authors show that for impulse noise corruption, differences between $\Loneloss$ and $\Ltwoloss$ are small and only visible for very high noise rates. 

In the case of experimental cryo-EM images, $\FRCloss$ loss outperforms $\Ltwoloss$  in the  FRC metrics. 
For this data set, which has only about 10,000  image pairs used during training, we see very high noise levels which affect the absolute FRC, MSE, and SSIM levels and obscure MSE and SSIM values. In the FRC plots we observe effects of over-fitting which result in the decay of the metric (calculated on 200 test set images, bottom row in Figure \ref{loss_training}). Interestingly, the over-fitting happens faster for the $\FRCloss$. We note that $\FRCloss$ is computationally more expensive per epoch; with the current implementation, training with the same number of epochs is about 1.5 times longer compared to $\Ltwoloss$, but the gains due to the loss properties (much fewer epochs needed to converge) are much larger. We also note that $\FRCloss$ does not optimise the absolute values of the image power-spectra, which can possibly drift during training.

\subsection{\bf{Denoising and signal power distribution}} \label{spectrum}

Previous studies have observed that denoising networks have an inherent spectral bias \cite{rahaman2018spectral, heckel2019denoising}, that skews their learning towards low complexity solutions. Practically, this manifests as a more robust learning in the lower frequencies.  Related work by \cite{ulyanov2018deep} reports a strong denoising ability on the low-frequencies when training a convolutional auto-encoder on a single noisy image and regularizing by early stopping. 
Employing FRC analysis we also observe that denoising low frequency is faster than denoising high frequency signal for both $\Ltwoloss$  and $\FRCloss$ (see Figure S2). 
In general, what these works (\cite{rahaman2018spectral, heckel2019denoising}) emphasize is that an over-parameterized network, where the network has sufficiently many parameters to represent an arbitrary image (including noise) perfectly, fits the signal of a natural image faster than noise. Furthermore, the very network itself has an inherent bias toward lower frequencies, meaning the signal restoration at lower frequencies is significantly better than mid to high range frequencies.  Several components of neural networks have been attributed to be the cause of this spectral bias, for instance the optimization algorithm itself, ReLu functions \cite{ulyanov2018deep,rahaman2019spectral} and pooling layers. Here, we investigate this spectral bias by denoising natural images with an altered power spectrum where all frequencies have equal mean power; we refer to these images as "power-normalised". We are interested in whether or not the spectral bias of neural networks is a matter of inherent preference for low frequencies or if it is rather just a coincidental correlation caused by the location of the largest signal power, which for natural images happens at low frequencies \cite{Bialek1993}. 
For this, we power-normalised 1000 GOI images, as explained in the Supplementary material and shown in Figure S3. The set of power normalised and original images (with the added Gaussian noise) were used to train two N2N networks with $\Ltwoloss$ loss. We find that the signal is recovered much more uniformly as a function of frequency for the network trained on power-normalised images (Figure  S4A). This suggests that spectral bias is in fact not a bias as much in the frequency aspect, as it is for the location of signal power.
As a consequence, neural networks learn better to denoise stronger signals, i.e./ the low frequency features. However, when the signal is normalised, denoising efficiency ceases to be strongly dependent on frequency, as can be seen from the FRC analysis.

\section{\textbf{Discussion and conclusions}}

Here, we explore denoising through the lens of the Fourier Ring Correlation and the associated $\FRCloss$ loss/metric function. First, we establish the $\FRCloss$ to be an effective metric, with sensitivity comparable to and exceeding the more established SSIM, while offering the advantage of qualitatively assessing the signal quality as a function of frequency. Further, we compare and contrast several denoising methods, including  U-nets and the BM3D method, and their ability to denoise high-frequency signals in images. We then explore the extent to which $\FRCloss$ may be used more directly in denoising---as a loss function for neural networks trained by back-propagation---by considering three different types of noise corruptions and comparing to the statistically optimal metric for each case. We conclude that all loss functions ultimately converge to similar denoising level, but  that the  convergence rate for $\FRCloss$ is markedly faster than for $\Loneloss$ and $\Ltwoloss$, albeit computationally more expensive per epoch. In parallel, we provide some mathematical properties of the Fourier Ring Correlation function by deriving its limits at low and high noise and by proving that $\FRCloss$ is analytically differentiable, thus qualifying as a valid loss function.

Finally, we use the FRC to characterise some peculiarities of neural network denoisers:  We confirm the spectral bias mentioned in a multitude of prior studies, but find it to be a circumstantial effect, caused by the majority of the signal strength being found at low frequencies. That is, the frequency-dependent denoising power of a network seems to be a knock-on effect of the spectral distribution of the signal and not inherent to the network itself. In the Supplementary Material we explore some potential FRC limitations due to denoising bias.


{
    \small
    \bibliographystyle{ieee_fullname}
    \bibliography{bibliography}

\begin{thebibliography}{10}\itemsep=-1pt

\bibitem{BANTERLE2013363}
Niccolò Banterle, Khanh~Huy Bui, Edward~A. Lemke, and Martin Beck.
\newblock Fourier ring correlation as a resolution criterion for
  super-resolution microscopy.
\newblock {\em Journal of Structural Biology}, 183(3):363--367, 2013.

\bibitem{batson2019noise2self}
Joshua Batson and Loic Royer.
\newblock Noise2self: Blind denoising by self-supervision.
\newblock {\em arXiv preprint arXiv:1901.11365}, 2019.

\bibitem{Berberich2021}
Andreas Berberich, Andreas Kurz, Sebastian Reinhard, Torsten~Johann Paul,
  Paul~Ray Burd, Markus Sauer, and Philip Kollmannsberger.
\newblock Fourier ring correlation and anisotropic kernel density estimation
  improve deep learning based smlm reconstruction of microtubules.
\newblock {\em Frontiers in Bioinformatics}, 1:55, 2021.

\bibitem{Culley:2018hx}
Si{\^a}n Culley, David Albrecht, Caron Jacobs, Pedro~Matos Pereira, Christophe
  Leterrier, Jason Mercer, and Ricardo Henriques.
\newblock {Quantitative mapping and minimization of super-resolution optical
  imaging artifacts}.
\newblock {\em Nature Methods}, pages 1--10, Feb. 2018.

\bibitem{dabov2007image}
Kostadin Dabov, Alessandro Foi, Vladimir Katkovnik, and Karen Egiazarian.
\newblock Image denoising by sparse 3-d transform-domain collaborative
  filtering.
\newblock {\em IEEE Transactions on image processing}, 16(8):2080--2095, 2007.

\bibitem{dabov2009bm3d}
Kostadin Dabov, Alessandro Foi, Vladimir Katkovnik, and Karen Egiazarian.
\newblock Bm3d image denoising with shape-adaptive principal component
  analysis.
\newblock In {\em SPARS'09-Signal Processing with Adaptive Sparse Structured
  Representations}, 2009.

\bibitem{Descloux:2019kh}
A Descloux, K~S Gru{\ss}mayer, and A Radenovic.
\newblock {Parameter-free image resolution estimation based on decorrelation
  analysis}.
\newblock {\em Nature Methods}, pages 1--11, Aug. 2019.

\bibitem{eskicioglu1995image}
Ahmet~M Eskicioglu and Paul~S Fisher.
\newblock Image quality measures and their performance.
\newblock {\em IEEE Transactions on communications}, 43(12):2959--2965, 1995.

\bibitem{harauz1986exact}
George Harauz and Marin van Heel.
\newblock Exact filters for general geometry three dimensional reconstruction.
\newblock {\em Optik (Stuttgart)}, 73(4):146--156, 1986.

\bibitem{heckel2019denoising}
Reinhard Heckel and Mahdi Soltanolkotabi.
\newblock Denoising and regularization via exploiting the structural bias of
  convolutional generators.
\newblock {\em arXiv preprint arXiv:1910.14634}, 2019.

\bibitem{koho2019fourier}
Sami Koho, Giorgio Tortarolo, Marco Castello, Takahiro Deguchi, Alberto
  Diaspro, and Giuseppe Vicidomini.
\newblock Fourier ring correlation simplifies image restoration in fluorescence
  microscopy.
\newblock {\em Nature communications}, 10(1):1--9, 2019.

\bibitem{krull2019noise2void}
Alexander Krull, Tim-Oliver Buchholz, and Florian Jug.
\newblock Noise2void-learning denoising from single noisy images.
\newblock In {\em Proceedings of the IEEE Conference on Computer Vision and
  Pattern Recognition}, pages 2129--2137, 2019.

\bibitem{krull2019probabilistic}
Alexander Krull, Tomas Vicar, and Florian Jug.
\newblock Probabilistic noise2void: Unsupervised content-aware denoising.
\newblock {\em arXiv preprint arXiv:1906.00651}, 2019.

\bibitem{OpenImages}
Alina Kuznetsova, Hassan Rom, Neil Alldrin, Jasper Uijlings, Ivan Krasin, Jordi
  Pont-Tuset, Shahab Kamali, Stefan Popov, Matteo Malloci, Alexander
  Kolesnikov, Tom Duerig, and Vittorio Ferrari.
\newblock The open images dataset v4: Unified image classification, object
  detection, and visual relationship detection at scale.
\newblock {\em IJCV}, 2020.

\bibitem{laine2019high}
Samuli Laine, Tero Karras, Jaakko Lehtinen, and Timo Aila.
\newblock High-quality self-supervised deep image denoising.
\newblock In {\em Advances in Neural Information Processing Systems}, pages
  6968--6978, 2019.

\bibitem{lehtinen2018noise2noise}
Jaakko Lehtinen, Jacob Munkberg, Jon Hasselgren, Samuli Laine, Tero Karras,
  Miika Aittala, and Timo Aila.
\newblock Noise2noise: Learning image restoration without clean data.
\newblock {\em arXiv preprint arXiv:1803.04189}, 2018.

\bibitem{kadid10k}
Hanhe Lin, Vlad Hosu, and Dietmar Saupe.
\newblock Kadid-10k: A large-scale artificially distorted iqa database.
\newblock In {\em 2019 Tenth International Conference on Quality of Multimedia
  Experience (QoMEX)}, pages 1--3. IEEE, 2019.

\bibitem{prakash2021fully}
Mangal Prakash, Alexander Krull, and Florian Jug.
\newblock Fully unsupervised diversity denoising with convolutional variational
  autoencoders.
\newblock In {\em International Conference on Learning Representations}, 2021.

\bibitem{prakash2020fully}
Mangal Prakash, Manan Lalit, Pavel Tomancak, Alexander Krul, and Florian Jug.
\newblock Fully unsupervised probabilistic noise2void.
\newblock In {\em 2020 IEEE 17th International Symposium on Biomedical Imaging
  (ISBI)}, pages 154--158. IEEE, 2020.

\bibitem{rahaman2019spectral}
Nasim Rahaman, Aristide Baratin, Devansh Arpit, Felix Draxler, Min Lin, Fred
  Hamprecht, Yoshua Bengio, and Aaron Courville.
\newblock On the spectral bias of neural networks.
\newblock In {\em International Conference on Machine Learning}, pages
  5301--5310. PMLR, 2019.

\bibitem{rahaman2018spectral}
Nasim Rahaman, Aristide Baratin, Devansh Arpit, Felix Draxler, Min Lin, Fred~A
  Hamprecht, Yoshua Bengio, and Aaron Courville.
\newblock On the spectral bias of neural networks.
\newblock {\em arXiv preprint arXiv:1806.08734}, 2018.

\bibitem{Bialek1993}
Daniel~L Ruderman and William Bialek.
\newblock Statistics of natural images: Scaling in the woods.
\newblock {\em Physical review letters}, 73(6):814, 1994.

\bibitem{SAXTON:1982jx}
W~O SAXTON and W BAUMEISTER.
\newblock {The Correlation Averaging of a Regularly Arranged Bacterial-Cell
  Envelope Protein}.
\newblock {\em Journal of Microscopy}, 127(AUG):127--138, 1982.

\bibitem{tian2020image}
Chunwei Tian, Yong Xu, and Wangmeng Zuo.
\newblock Image denoising using deep cnn with batch renormalization.
\newblock {\em Neural Networks}, 121:461--473, 2020.

\bibitem{ulyanov2018deep}
Dmitry Ulyanov, Andrea Vedaldi, and Victor Lempitsky.
\newblock Deep image prior.
\newblock In {\em Proceedings of the IEEE Conference on Computer Vision and
  Pattern Recognition}, pages 9446--9454, 2018.

\bibitem{van1987similarity}
Marin Van~Heel.
\newblock Similarity measures between images.
\newblock {\em Ultramicroscopy}, 21(1):95--100, 1987.

\bibitem{vanHeel1982}
M Van~Heel, W Keegstra, W Schutter, and EJF Van~Bruggen.
\newblock Arthropod hemocyanin structures studied by image analysis.
\newblock {\em Life Chem. Rep. Suppl}, 1:69--73, 1982.

\bibitem{van2005fourier}
Marin Van~Heel and Michael Schatz.
\newblock Fourier shell correlation threshold criteria.
\newblock {\em Journal of structural biology}, 151(3):250--262, 2005.

\bibitem{wang2004image}
Zhou Wang, Alan~C Bovik, Hamid~R Sheikh, and Eero~P Simoncelli.
\newblock Image quality assessment: from error visibility to structural
  similarity.
\newblock {\em IEEE transactions on image processing}, 13(4):600--612, 2004.

\bibitem{wang2003multiscale}
Zhou Wang, Eero~P Simoncelli, and Alan~C Bovik.
\newblock Multiscale structural similarity for image quality assessment.
\newblock In {\em The Thrity-Seventh Asilomar Conference on Signals, Systems \&
  Computers, 2003}, volume~2, pages 1398--1402. Ieee, 2003.

\bibitem{weigert2018content}
Martin Weigert, Uwe Schmidt, Tobias Boothe, Andreas M{\"u}ller, Alexandr
  Dibrov, Akanksha Jain, Benjamin Wilhelm, Deborah Schmidt, Coleman Broaddus,
  Si{\^a}n Culley, et~al.
\newblock Content-aware image restoration: pushing the limits of fluorescence
  microscopy.
\newblock {\em Nature methods}, 15(12):1090--1097, 2018.

\bibitem{BM3DNetAC}
Dong Yang and J. Sun.
\newblock Bm3d-net: A convolutional neural network for transform-domain
  collaborative filtering.
\newblock {\em IEEE Signal Processing Letters}, 25:55--59, 2018.

\bibitem{zhang2017beyond}
Kai Zhang, Wangmeng Zuo, Yunjin Chen, Deyu Meng, and Lei Zhang.
\newblock Beyond a gaussian denoiser: Residual learning of deep cnn for image
  denoising.
\newblock {\em IEEE Transactions on Image Processing}, 26(7):3142--3155, 2017.

\bibitem{zhang2018ffdnet}
Kai Zhang, Wangmeng Zuo, and Lei Zhang.
\newblock Ffdnet: Toward a fast and flexible solution for cnn-based image
  denoising.
\newblock {\em IEEE Transactions on Image Processing}, 27(9):4608--4622, 2018.

\bibitem{zhao2015loss}
Hang Zhao, Orazio Gallo, Iuri Frosio, and Jan Kautz.
\newblock Loss functions for neural networks for image processing.
\newblock {\em arXiv preprint arXiv:1511.08861}, 2015.

\end{thebibliography}


\begin{thebibliography}{1}\itemsep=-1pt

\bibitem{OpenImages}
Alina Kuznetsova, Hassan Rom, Neil Alldrin, Jasper Uijlings, Ivan Krasin, Jordi
  Pont-Tuset, Shahab Kamali, Stefan Popov, Matteo Malloci, Alexander
  Kolesnikov, Tom Duerig, and Vittorio Ferrari.
\newblock The open images dataset v4: Unified image classification, object
  detection, and visual relationship detection at scale.
\newblock {\em IJCV}, 2020.

\bibitem{lehtinen2018noise2noise}
Jaakko Lehtinen, Jacob Munkberg, Jon Hasselgren, Samuli Laine, Tero Karras,
  Miika Aittala, and Timo Aila.
\newblock Noise2noise: Learning image restoration without clean data.
\newblock {\em arXiv preprint arXiv:1803.04189}, 2018.

\bibitem{kadid10k}
Hanhe Lin, Vlad Hosu, and Dietmar Saupe.
\newblock Kadid-10k: A large-scale artificially distorted iqa database.
\newblock In {\em 2019 Tenth International Conference on Quality of Multimedia
  Experience (QoMEX)}, pages 1--3. IEEE, 2019.

\bibitem{Miura2021}
Kota Miura and Simon~F Nørrelykke.
\newblock Reproducible image handling and analysis.
\newblock {\em The EMBO Journal}, 40(3):e105889, 2021.

\bibitem{palovcak2020enhancing}
Eugene Palovcak, Daniel Asarnow, Melody~G Campbell, Zanlin Yu, and Yifan Cheng.
\newblock Enhancing snr and generating contrast for cryo-em images with
  convolutional neural networks.
\newblock {\em bioRxiv}, 2020.

\bibitem{ronneberger2015u}
Olaf Ronneberger, Philipp Fischer, and Thomas Brox.
\newblock U-net: Convolutional networks for biomedical image segmentation.
\newblock In {\em International Conference on Medical image computing and
  computer-assisted intervention}, pages 234--241. Springer, 2015.

\bibitem{scheres2012relion}
Sjors~HW Scheres.
\newblock Relion: implementation of a bayesian approach to cryo-em structure
  determination.
\newblock {\em Journal of structural biology}, 180(3):519--530, 2012.

\end{thebibliography}
}

\end{document}


\maketitle

\section{\textbf{Supplementary Methods}} \label{methods}

\subsection{\bf{Data sets}} \label{data_pipeline}

We worked with Google Open Images (GOI) \cite{OpenImages}, KADID-10 \cite{kadid10k}, and an experimental Cryo-EM image dataset from the Relion tutorial (\cite{scheres2012relion}, \url{ftp://ftp.mrc-lmb.cam.ac.uk/pub/scheres/relion30_tutorial_data.tar}, part of the EMPIAR-10204 data set). Fifty-thousand images were selected from the GOI set for training of networks. Fifty random GOI images were used for tests and analysis. For training, random image crops, of size 256x256 pixels, were used. For testing, images were cropped to a square and resized to the size of 256x256. The color channels were averaged. The pixel values were  normalized to the range -0.5 to 0.5.
We used all 81 images from the KADID-10 data set and four selected distortion types (Gaussian and impulse noise, jitter and motion blur). 

The experimental Cryo-EM image dataset was processed to extract about 10'000 images of proteasome protein particles. For each particle, two noisy realisations were computed from the original movie stack by averaging 35 frames for each output, of size 256x256.

The Gaussian noise corruptions were zero-mean centered and produced with a standard deviation set to 0.4. The log-normal corruptions were zero-median centered with a standard deviation set to 1.3. 

To construct the power-normalised data set we selected 1000 random images from the GOI data set. The same set of images was used to train the $\Ltwoloss$-N2N-GOI1000 network shown in Figure 2 (grey dashed line, Gaussian noise corruptions with standard deviation set to 0.4)   The power spectrum of images was normalised as described in Figure \ref{s3}. Gaussian noise with standard deviation of 0.05 was used to produce corruptions. 

\subsection{\bf{Neural network training and denoising details}} \label{training} 
We showcase our work on one of the most used network architectures, the U-net\cite{ronneberger2015u}, with the Noise2Noise training scheme \cite{lehtinen2018noise2noise}. All networks were trained for 200 epochs, with a minibatch of size 10, learning rate 0.0003, ramp down rate of 0.3 and Adam optimiser. Noise was generated every epoch during training, except for the EM data set, where no additional noise was added to the input.

We used the BM3D algorithm from the Python pip package, with $\mathrm{stage\_arg}$ set to $ALL\_STAGES$ argument. A Gaussian filter from the scikit-image package was used with sigma=1 and mode='wrap'. Hann window function from the scikit-image package with default parameters was used.

\subsection{Code availability}
Our source code is publicly available at \url{https://github.com/frcCVPR/frc-loss}.

\section{\bf{Supplementary results}} \label{bias}

\subsection{\bf{Denoising and bias}}

A typical scientific image processing pipeline is a multistep process \cite{Miura2021}, where the individual steps consist of the image data being normalized, aligned, averaged, illumination corrected, filtered, segmented, etc. It is not a priori clear how image denoising affects such processing or exactly at which step in the pipeline it should take place. 
For instance, in single-particle cryo-EM experiment, images acquired with the electron microscope are sometimes averaged several thousand times in-between rounds of classification and image registration \cite{scheres2012relion}. This elaborate pipeline is a response to the extremely low SNR of single particle cryo-EM data. 

Employing FRC analysis we test the usability of denoising as part of a post-processing routine that executes image averaging. We use the $\Ltwoloss$-N2N network trained with Gaussian noise (orange line in Figure 2), to denoise an image with 200 independent noise realizations and subsequently average these individually denoised images. 
We observe that denoising the same signal, with independent noise realizations, and further averaging yields only limited improvement. Furthermore, the signal can be better restored by direct averaging rather than denoising and averaging the individual noisy images (see FRC analysis in Figure \ref{s4}B). 
A denoised image $D$ consists of several components, $D= S + N_D + B$, with signal $S$, remaining noise after denoising $N_{D}$, and denoising bias $B$ (see detailed analysis in \cite{palovcak2020enhancing}). 
The denoising bias $B$ is pronounced for high frequencies and cannot be removed by averaging the individually denoised images from independent noise realisations. 
These results suggest that for scenarios requiring extensive image averaging, caution should be exercised when applying network based denoising.
Possibly, networks optimised for bias $B$ could perform better than the one we explored here, or perhaps denoising should be applied only as the very last processing step.
That the outcome differs when swapping the order of the two operations (averaging and network-denoising) should not surprise, as the network operation on an image is not a linear operation, and our results here highlight and exemplify this noncommutative property. Clearly, more work is required to better understand and minimize this denoising-introduced bias.

\clearpage
\section{\textbf{Supplementary Figures}} 
\renewcommand{\thefigure}{S\arabic{figure}}
\begin{figure}[h]
{\centering
\onecolumn\includegraphics[width=0.8\textwidth]{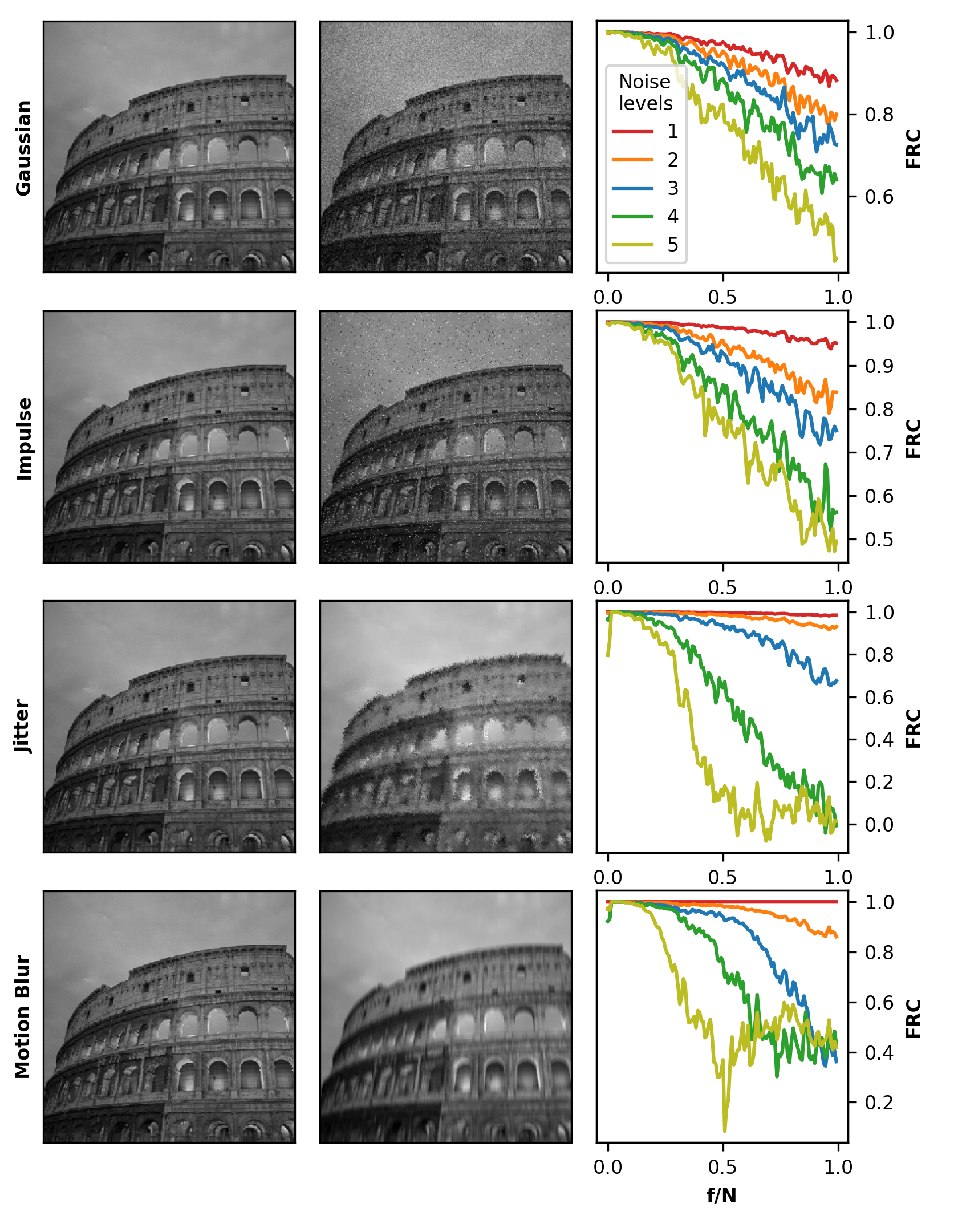}
\caption{FRC curves for distorted against ground truth images for four different corruption types (Gaussian and impulse noise, jitter and motion blur) and five corruption levels (plots in the right column). Highest corruption (level 5)  images, selected from the Konstanz artificially distorted image quality database KADID-10k, are shown in the middle column, together with lowest corruption (level 1) images in the left column,  $f/N$ denotes frequency normalised by the Nyquist frequency for the images.}
\label{corruption}}
\footnotesize
\vspace{\baselineskip}

\end{figure}

\begin{figure*}[p]
\centering
\includegraphics[width=0.9\textwidth]{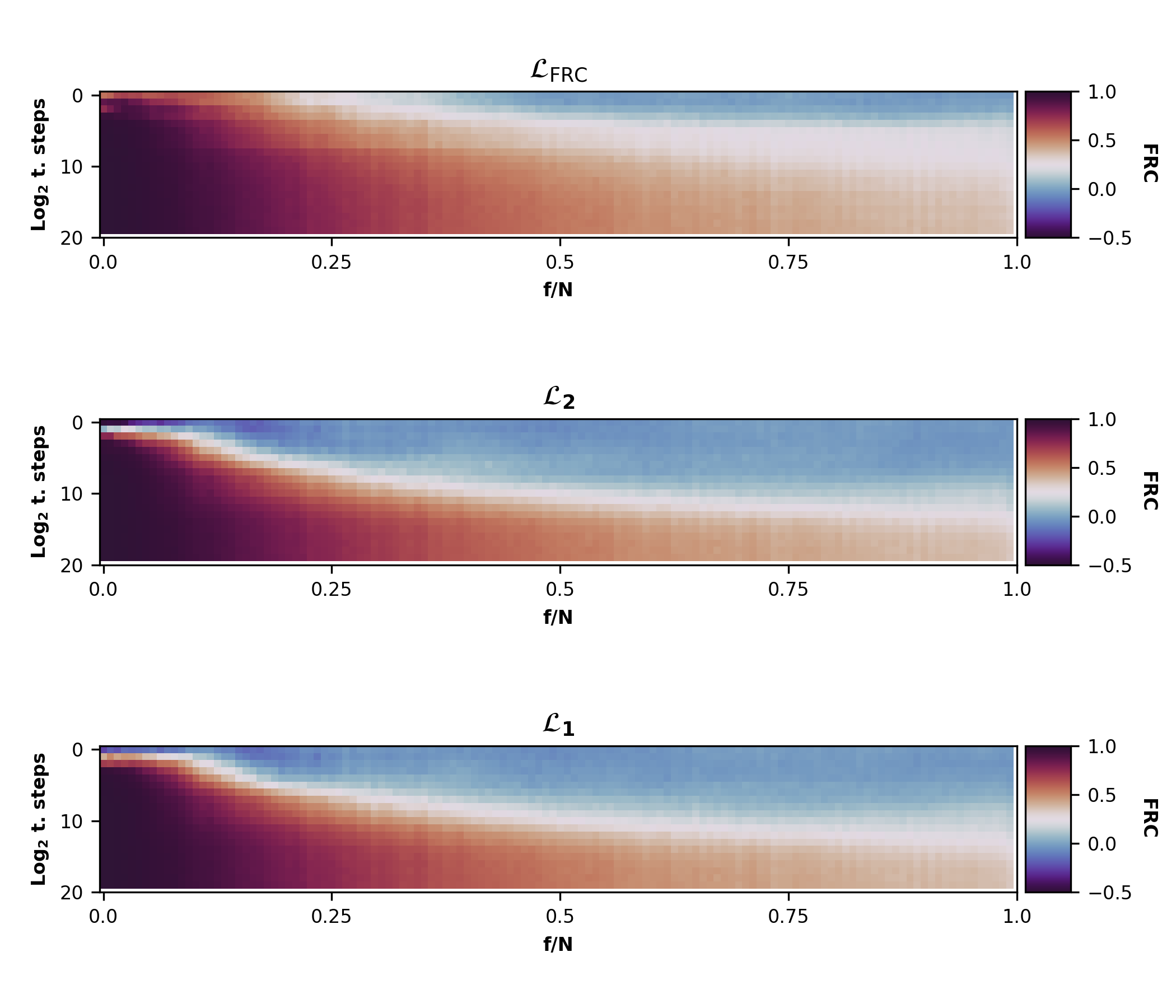}
\caption{Low frequencies are denoised faster than high frequencies. FRC is shown as a function of training steps (each step is one processed batch) for the full frequency spectrum. The top row histogram shows training with the $\FRCloss$, the middle row with $\Ltwoloss$, and the bottom row with $\Loneloss$. The data corresponds to the data shown in Figure 4, top row. The results are calculated as averages for the 50 test images from the GOI.} 
\label{figure_s1}
\end{figure*}

\begin{figure*}[p]
\centering
\includegraphics[width=\textwidth]{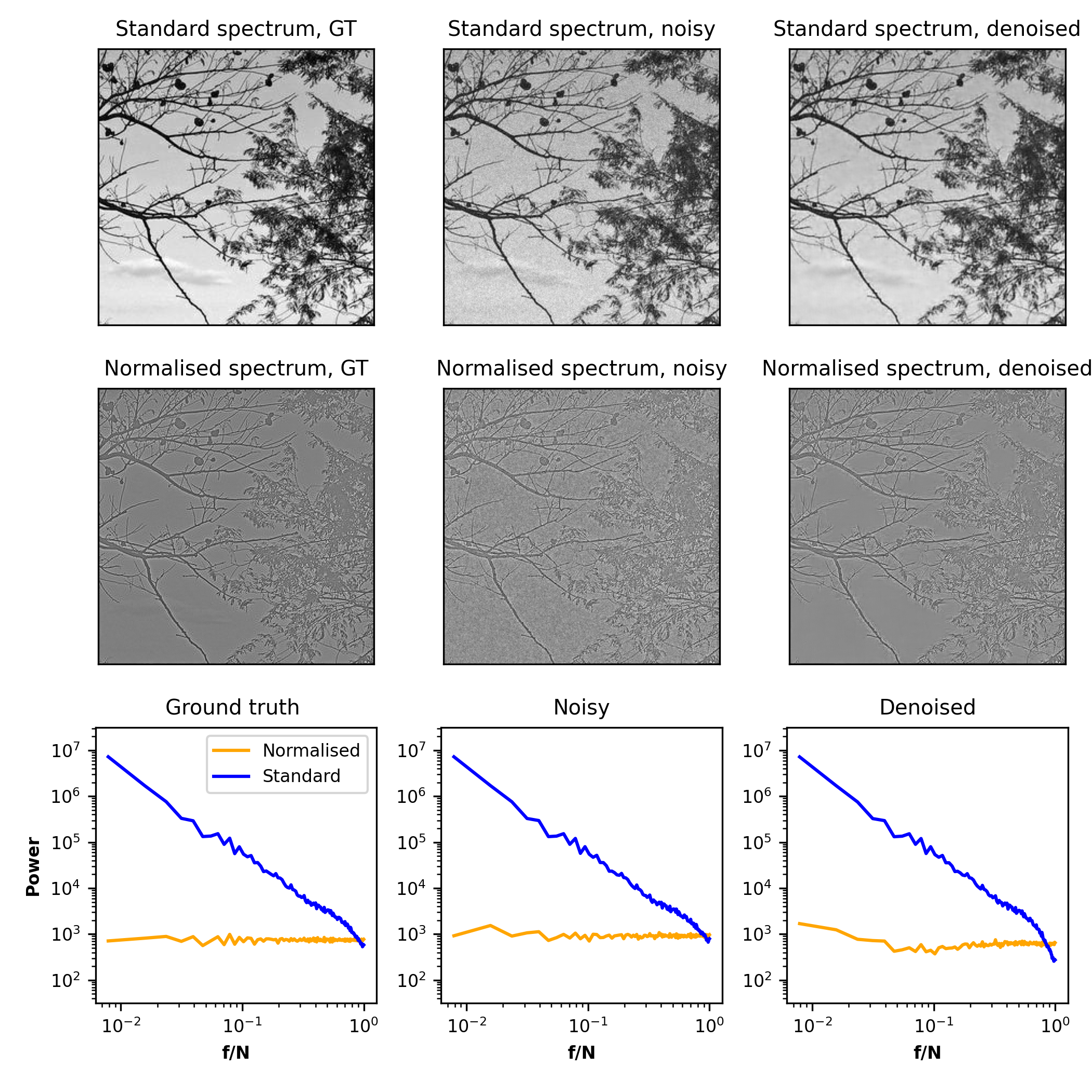}
\caption{Examples of a natural image with a normalised power spectrum (and the same total power as in the original image). Top row: the original, noisy and denoised image with a normal power spectrum. Middle row: Same image but with a normalised power spectrum. Bottom row: The power spectra corresponding to each column. }
\label{s3}
\end{figure*}

\begin{figure*}[p]
\centering
\includegraphics[width=0.5\textwidth]{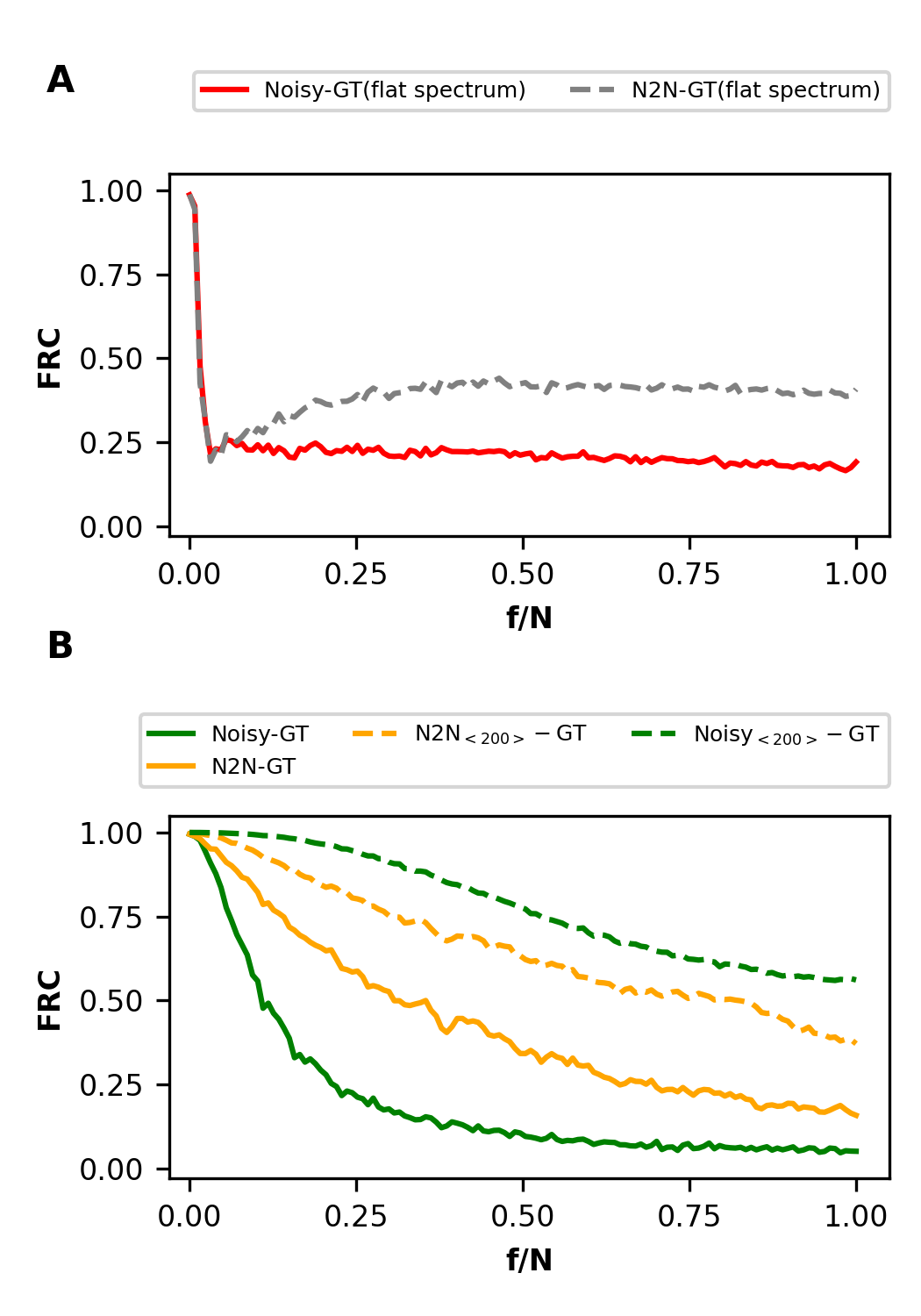}
\caption{$\mathbf{A}$ Denoising performance for normalised power spectrum images. The FRC curves with respect to the ground truth, power spectrum normalised, images  were obtained by averaging individual FRC curves for 50 denoised (grey-dashed line) or noisy (red line) training images. Pre-processing with a Hann window to suppress edge artifacts in the Fourier signal was performed before FRC calculation.  The network was trained with $\Ltwoloss$ using N2N scheme on 1000 GOI power spectrum normalised images with a noise standard deviation adjusted to result in FRC to GT of around 0.25 (red line). $\mathbf{B}$. FRC curves for denoised and post-processed,  i.e. further averaged, images. Green line: noisy image vs GT; orange line: denoised image vs GT; green dashed line: average image from 200 noisy image realisations vs GT, orange dashed line: average image from 200 denoised images vs GT. Each line is reported as an average over set of 50 test images from GOI. Pre-processing with a Hann window to suppress edge artifacts in the Fourier signal was performed before FRC calculation. For denoising $\Ltwoloss$-N2N network trained on the GOI data set (see Figure 2) was used. 
\label{s4}
}
\end{figure*}

\twocolumn

\renewcommand{\refname}{Supplementary References}
{
    \clearpage
    \small
    \bibliographystyle{ieee_fullname}
    \bibliography{bibliography}
}